%% file: Doty.tex
\documentclass[proceedings]{stacs}
\stacsheading{2010}{275-286}{Nancy, France}
\firstpageno{275}

\usepackage{amsmath}
\usepackage{amssymb}
\usepackage{ifpdf}
\usepackage{algorithmic}
\usepackage{subfig}
\usepackage{cite}
\usepackage{graphicx}
\usepackage{rotating}

    \usepackage{commands-tam}
    \usepackage{amsfonts}
    \usepackage{amsthm}

\newcommand{\calS}{\mathcal{S}}

\providecommand{\calT}{\mathcal{T}}

\ifpdf

  \usepackage[pdftex]{epsfig}
  \opt{normal,focs}{\usepackage[pdftex]{hyperref}}

\else
    \usepackage[dvips]{epsfig}
    \newcommand{\href}[2]{#2}

\fi

\newcommand{\REPR}{\mathsf{REPR}}
\newcommand{\FIN}{\mathsf{FIN}}

\begin{document}

\title[Intrinsic Universality in Self-Assembly]{Intrinsic Universality in Self-Assembly}

\author[lab2]{D. Doty}{David Doty}
\address[lab2]{Department of Computer Science, University of Western Ontario \newline London, Ontario, Canada, N6A5B7.} 
\email[David Doty]{ddoty@csd.uwo.ca}
\author[lab1]{J.H. Lutz}{Jack H. Lutz}
\author[lab1]{M.J. Patitz}{Matthew J. Patitz}
\author[lab1]{S.M. Summers}{Scott M. Summers}
\address[lab1]{Department of Computer Science, Iowa State University
  \newline Ames, IA 50011 USA.}  
\email[Jack H. Lutz]{lutz@cs.iastate.edu} 
\email[Matthew J. Patitz]{patitz@cs.iastate.edu}
\email[Scott M. Summers]{summers@cs.iastate.edu}  
\author[lab5]{D. Woods}{Damien Woods}
\address[lab5]{California Institute of Technology, Pasadena, CA 91125, USA.}	 
\email{woods@caltech.edu}  

\thanks{This research was supported in part by National Science Foundation
Grants 0652569, 0728806 and 0832824, by the Spanish Ministry of Education
and Science (MEC) and the European Regional Development Fund (ERDF) under
project TIN2005-08832-C03-02, by Junta de Andaluc\'{i}a grant TIC-581,
and by Natural Sciences and Engineering Research Council of Canada (NSERC)
Discovery Grant R2824A01 and the Canada Research Chair Award in Biocomputing.} 

\subjclass{Theory}


\begin{abstract}
\noindent We show that the Tile Assembly Model exhibits a strong notion of universality where the goal is to give a single tile assembly system that simulates the behavior of any other tile assembly system. We give a tile assembly system that is capable of simulating a very wide class of tile systems, including itself. Specifically, we give a tile set that simulates the assembly of any tile assembly system in a class of systems that we call \emph{locally consistent}: each tile binds with exactly the strength needed to stay attached, and that there are no glue mismatches between tiles in any produced assembly.



  Our construction is reminiscent of the studies of \emph{intrinsic universality} of cellular automata by Ollinger and others, in the sense that our simulation of a tile system $T$ by a tile system $U$ represents each tile in an assembly produced by $T$ by a $c \times c$ block of tiles in $U$, where $c$ is a constant depending on $T$ but not on the size of the assembly $T$ produces (which may in fact be infinite). Also, our construction improves on earlier simulations of tile assembly systems by other tile assembly systems (in particular, those of Soloveichik and Winfree, and of Demaine et al.) in that we simulate the actual process of self-assembly, not just the end result, as in Soloveichik and Winfree's construction, and we do not discriminate against infinite structures. Both previous results simulate only temperature 1 systems, whereas our construction simulates tile assembly systems operating at temperature 2.
\end{abstract}

\maketitle

\input{section1}

\input{section2}

\input{section3}
\bibliographystyle{amsplain}
\bibliography{main,dim,random,dimrelated,rbm,tam,ca}


\end{document}

%% file: section1.tex
\section{Introduction}

The development of DNA tile self-assembly has moved nanotechnology
closer to the goal of engineering useful systems that assemble
themselves from molecular components.  Since Seeman's pioneering
work in the 1980s \cite{Seem82}, many laboratory experiments have
shown that DNA tiles can be designed to spontaneously assemble
with one another into desired structures \cite{RoPaWi04}.  As
physical and mathematical error-suppression techniques improve
\cite{MLR07,ChenGoel04,SolWin05,
MajSahLaBRei06,FujZhaWinMur08}, this molecular programming of matter will become
practical at ever larger scales.

The Tile Assembly Model, developed by Winfree \cite{Winf98,RotWin00}, is a discrete
mathematical model of DNA tile self-assembly that enables us to
explore the potentialities and limitations of this kind of
molecular programming. It is essentially an ``effectivization'' of classical Wang tiling \cite{Wang61} in which the fundamental components are un-rotatable, but translatable square ``tile types'' whose sides are labeled with glue ``colors'' and ``strengths.'' Two tiles that are placed next to each other \emph{interact} if the glue colors on their abutting sides
match, and they \emph{bind} if the strength on their abutting sides matches with total strength at least a certain ambient ``temperature.'' Extensive refinements of the abstract Tile Assembly Model were given by Rothemund and Winfree in \cite{RotWin00,Roth01}.  (Consult the technical appendix for full details of the abstract Tile Assembly Model.) The model deliberately oversimplifies the physical realities of self-assembly, but Winfree proved that it is
Turing universal \cite{Winf98}, implying that
self-assembly can be algorithmically directed.

In this paper we investigate whether the Tile Assembly Model is capable of a much
stronger notion of universality where the goal is to give a single tile assembly system that simulates the behavior of any other tile assembly system. We give a tile assembly system that is capable of simulating a very wide class of
tile systems, including itself. Our notion of simulation is inspired by, but somewhat stronger than, intrinsic universality in cellular automata~\cite{Ollinger-CSP08,Ollinger-STACS03,AlbertCulik87,DurandRoka89,Margenstern07}. In our construction a simulated tile assembly system is encoded in a seed assembly of the simulating system. This encoding is done in a very simple (logspace computable) way. The seed assembly then grows to form an assembly that is a re-scaled (larger) version of the simulated assembly, where each tile in the latter is represented by a supertile (square of tiles) in the simulator. Not only this, but each of the possible (nondeterministically chosen) assembly sequences of the simulated tile system is modeled by a possible assembly sequence in the simulating system (also nondeterministically chosen). The latter property of our system is important and highlights one way in which this work distinguishes itself from other notions of intrinsic universality found in the cellular automata literature: not only do we want to simulate the final assembly but we also want the simulator to have the ability to dynamically simulate each of the valid growth processes that could lead to that final assembly.

A second distinguishing property of our universal tile set is that it simulates nondeterministic choice in a ``fair'' way.
An inherent feature of the Tile Assembly Model is the fact there are often multiple (say $k$) tiles that can go into any one position in an assembly sequence, and one of these $k$ is nondeterministically chosen. One way to simulate this feature is to nondeterministically choose which of $k$ supertiles should grow in the analogous (simulated) position. However, due to the size blowup in supertiles caused by encoding an arbitrary-sized simulated tile set into a fixed-sized universal tile set, it seems that we need to simulate one nondeterministic choice by using a sequence of nondeterministic choices within the supertile. Interpreting the nondeterministic choice to be made according to uniform random selection, if the selection by the simulating tile set is implemented in a na\"{\i}ve way, this can lead to unfair selection: when selecting 1 supertile out of $k$, some supertiles are selected with extremely low probability. To get around this problem, our system uses a random number selector that chooses a random tile with probability $\Theta({1}/{k})$ and so we claim that we are simulating nondeterminism in a ``fair'' way.

Thirdly, the Tile Assembly Model has certain geometric constraints that are not seen in cellular automata, and this adds some difficulty to our construction. Existing techniques for constructing intrinsically universal cellular automata are not directly applicable to tile assembly. For example, when a tile is placed at a position, that position can not be reused for further ``computation'' and this presents substantial difficulties when trying to fit the various components of our construction into a supertile. Each supertile encodes the entire simulated tile set and has the functionality to propagate this information to other (yet to be formed) supertiles. Not only this, each supertile must decide which tile placement to simulate, whilst making (fair) nondeterministic choices if necessary. Finally, each supertile should correctly propagate (output) sides that are consistent with the chosen supertile. We give a number of figures to illustrate how these goals were met within the geometric constraints of the model.

Our main result presented in this paper is, in some sense, a
continuation of some previous results in self-assembly. For instance,
Soloveichik and Winfree \cite{SolWin07} exhibit a beautiful connection
between the Kolmogorov complexity of a finite shape $X$ and the minimum
number of tiles types needed to assemble $X$. It turns out that their
construction can be made to be ``universal'' in the following sense: there exists a tile set $T$, such that for every
``temperature 1'' tile assembly system that produces a finite shape whose underlying binding graph is a spanning tree, $T$ simulates the given temperature 1 tile system with a
corresponding blow-up in the scale. Note that this method restricts
the simulated tile system to be temperature 1, i.e., a non-cooperative
tile assembly system, which are conjectured \cite{LSAT1} to produce
``simple'' shapes and patterns in the sense of Presburger
arithmetic \cite{Presburger30}.

A similar result, recently discovered by Demaine, Demaine, Fekete, Ishaque, Rafalin, Schweller, and Souvaine
\cite{DDFIRSS07}, established the existence of a
general-purpose ``staged-assembly'' system that is capable of simulating
any temperature 1 tile assembly system that produces a ``fully
connected'' finite shape. Note that, in this construction, the scaling
factor is proportional to $O(\log |T|)$, where $T$ is the simulated tile
set. This construction has the desirable property that the set of tile
types belonging to the simulator is general purpose (i.e., the size of the simulator tile set is independent of the to-be-simulated tile set) and all of the information
needed to carry out the simulation is, in some sense, encoded in a
sequence of laboratory steps. An open question in
\cite{DDFIRSS07} is whether or not their construction can
be augmented to handle temperature 2 tile assembly systems.

Our construction is general enough to be able to simulate powerful and interesting tile sets, yet sufficiently simple so that it actually belongs to the class of tile assembly systems that it can simulate, a class we term \emph{locally consistent}. Systems in this class have the properties that each tile binds with exactly strength 2, and there are no glue mismatches in any producible assembly. This captures a wide class of tile assembly systems, including counters, square-builders and other shape-building tile assembly systems, and the tile assembly systems described in \cite{AdChGoHu01,RotWin00,SolWin07,CCSA}.
Modulo re-scaling, our universal tile set can be said to display the characteristics of the entire collection of tile sets in its class. Our construction is a {\em direct simulation} in that the technique does not involve the simulation of intermediate models (such as circuits or Turing machines), which have been used in intrinsically universal cellular automata constructions~\cite{Ollinger-CSP08}.

One of the nice properties of intrinsic universality~\cite{Ollinger-CSP08} is that it provides a clear definition that facilitates proofs that a given tile set is not universal. We leave as an open problem the intrinsic universality status of the Tile Assembly Model in its full generality.

Lafitte and Weiss \cite{LafitteW07,LafitteW08,LafitteW09} have also studied universality in the related model of
Wang tiling~\cite{Wang61}. Some of their definitions, particularly in
\cite{LafitteW08}, are similar to our definitions of
simulation and universality, and also to those of Ollinger \cite{Ollinger-CSP08}.
However, Wang tiling is not a model of self-assembly, as it is concerned
with the ability of finite tile sets to tile the whole plane (with no
mismatches), without regard to the \emph{process} by which these tiles are
placed.  What is important is simply the existence of some valid
tiling.  In the TAM, which takes the order in
which tiles are placed, one by one, into account, it must be shown that
not only is there a sequence by which tiles could be individually and
stably added to form the output assembly, but that \emph{every} possible
such sequence leads to the desired output.  Furthermore, in the TAM a
tile addition can be valid even if it causes mismatches as long as it is stable.

Most attempts to adapt the constructions of Wang tiling studies (such as those
in \cite{LafitteW07,LafitteW08,LafitteW09}) to self-assembly result in a tile assembly system in
which many junk assemblies are formed due to incorrect nondeterministic
choices being made that arrest any further growth and/or result in
assemblies which are inconsistent with the desired output assembly.
We therefore require novel techniques to ensure that no nondeterminism is
introduced, other than that already present in the tile system being simulated,
and that the only produced assemblies are those that represent the intended
result or valid partial progress toward it.

%% file: section2.tex
\section{Intrinsic Universality in Self-Assembly}
\newcommand{\frakC}{\mathfrak{C}}

In this section, we define our notion of intrinsic universality of tile assembly systems. It is inspired by, but distinct from, similar notions for cellular automata \cite{Ollinger-CSP08}. Where appropriate, we identify where some part of our definition differs from the ``corresponding'' parts in \cite{Ollinger-CSP08}, typically due to a fundamental difference between the abstract Tile Assembly Model
and cellular automata models.

Intuitively, a tile set $U$ is universal for a class $\frakC$ of tile assembly systems if $U$ can ``simulate'' any tile assembly system in $\frakC$, where we use an appropriate seed assembly to give a tile assembly system~$\mathcal{U}$. $U$ is intrinsically universal if the simulation of~$\mathcal{T}$ by~$\mathcal{U}$ can be done according to a simple ``block substitution scheme'' where equal-size square blocks of tiles in assemblies produced by $\mathcal{U}$ represent tiles in assemblies produced by~$\mathcal{T}$. Furthermore, since we wish to simulate the entire process of self-assembly, and not only the final result, it is critical that the simulation be such that the ``local transition rules'' involving intermediate producible (and nonterminal) assemblies of $\mathcal{T}$ be faithfully represented in the simulation.

In the subsequent definitions, given two partial functions $f,g$, we write $f(x) = g(x)$ if~$f$ and~$g$ are both defined and equal on~$x$, or if~$f$ and~$g$ are both undefined on $x$. Let $c,c'\in\N$, let $[c:c']$ denote the set $\{c,c+1,\ldots,c'-1\}$, and let $[c]$ denote the set $[0:c] = \{0,1,\ldots,c-1\}$, so that $[c]^2$ forms a $c\times c$ square with the origin as the lower-left corner.

The natural analog of a configuration of a cellular automaton is an assembly of a tile assembly system. However, unlike cellular automata in which every cell has a well-defined state, in tile assembly, there is a fundamental difference between a point being empty space and being occupied by a tile. Therefore we keep the convention of representing an assembly as a partial function $\alpha:\Z^2 \dashrightarrow T$ (for some tile set $T$), rather than treating empty space as just another type of tile.

Let $\mathcal{T}=(T,\sigma_\mathcal{T},\tau)$ and $\calS=(S,\sigma_\calS,\tau)$ be tile assembly systems. For simplicity, assume that $\sigma_\mathcal{T}(0,0)$ is defined, and $\sigma_\mathcal{T}$ is undefined on $\Z^2 - \{(0,0)\}$ (i.e., $\mathcal{T}$ is singly-seeded with the seed tile placed at the origin). We will use this assumption of a single seed throughout the paper, but it is not strictly necessary and is only used for simplicity of discussion. Define a \emph{representation function} to be a partial function of the form $r:([c]^2 \dashrightarrow S) \dashrightarrow T$. That is, $r$ takes a pattern $p:[c]^2 \dashrightarrow S$ of tile types from~$S$ painted onto a $c \times c$ square (with locations at which $p$ is undefined representing empty space), and (if $r$ is defined for input $p$) gives a single tile type from $T$. Intuitively, $r$ tells us how to interpret $c \times c$ blocks within assemblies of $\calS$ as single tiles of $T$. We write $\REPR$ for the set of all representation functions.

We say $\calS$ \emph{(intrinsically) simulates $\mathcal{T}$ with resolution loss $c$} if there exists a representation function $r:([c]^2 \dashrightarrow S) \dashrightarrow T$ such that the following conditions hold.
\begin{enumerate}
  \item $\dom \sigma_\calS \subseteq [c]^2$ and $r(\sigma_\calS) = \sigma_{\mathcal{T}}(0,0)$, i.e., the seed assembly of $\calS$ represents the seed of $\mathcal{T}$.

  \item \label{simulate-cond-2} For every producible assembly $\alpha_\mathcal{T} \in \prodasm{\mathcal{T}}$ of $\mathcal{T}$, there is a producible assembly $\alpha_\calS \in \prodasm{\calS}$ of $\calS$ such that, for every $x,y\in\Z$,
      $$
      r \left(  \left( \alpha_\calS \upharpoonright ([cx:c(x+1)] \times [cy:c(y+1)]) \right) + (-c x, -c y)  \right) = \alpha_\mathcal{T}(x,y).
      $$
      That is, the $c \times c$ block at (relative) position $(x,y)$ (relative to the other $c \times c$ blocks; the absolute position is $(c x,c y)$) of assembly $\alpha_\calS$ represents the tile type at (absolute) position $(x,y)$ of assembly $\alpha_\mathcal{T}$. In this case, write $r^*(\alpha_\calS) = \alpha_\mathcal{T}$; i.e., $r$ induces a function $r^*:\prodasm{\calS} \to \prodasm{\mathcal{T}}$.

  \item For all $\alpha_\mathcal{T},\alpha_\mathcal{T}' \in \prodasm{\mathcal{T}}$, it holds that
      $\alpha_\mathcal{T} \to_\mathcal{T} \alpha_\mathcal{T}'$
      if and only if there exist $\alpha_\calS,\alpha_\calS' \in \prodasm{\calS}$ such that $r^*(\alpha_\calS) = \alpha_\mathcal{T}$, $r^*(\alpha_\calS') = \alpha_\mathcal{T}'$ (in the sense of condition \eqref{simulate-cond-2}), and $\alpha_\calS \to_\calS \alpha_\calS'$. That is, every valid assembly sequence of $\mathcal{T}$ can be ``mimicked'' by $\calS$, but no other assembly sequences can be so mimicked, so that the meaning of the relation $\to$ is preserved by $r^*$.
\end{enumerate}

Let $\frakC$ be a class of singly-seeded tile assembly systems, and let $U$ be a tile set (with tile assembly systems having tile set $U$ not necessarily elements of $\frakC$). Note that every element of $\frakC$, $\REPR$, and $\FIN(U)$ is a finite object, hence can be represented in a suitable format for computation in some formal system such as Turing machines. We say $U$ is \emph{(intrinsically) universal} for $\frakC$ if there are computable functions $R:\frakC \longrightarrow \REPR$ and $A:\frakC \longrightarrow \FIN(U)$ such that, for each $\mathcal{T} = (T,\sigma_\mathcal{T},\tau) \in \frakC$, there is a constant $c\in\N$ such that, letting $r = R(\mathcal{T})$, $\sigma=A(\mathcal{T})$, and $\mathcal{U}_\mathcal{T} = (U,\sigma,\tau)$, $\mathcal{U}_\mathcal{T}$ simulates $\mathcal{T}$ with resolution loss $c$ and representation function $r$. That is, $R(\mathcal{T})$ outputs a representation function that interprets assemblies of $\mathcal{U}_\mathcal{T}$ as assemblies of $\mathcal{T}$, and $A(\mathcal{T})$ outputs the seed assembly used to program tiles from $U$ to represent the seed tile of $\mathcal{T}$.

%% file: section3.tex
\section{An Intrinsically Universal Tile Set}

In this section, we exhibit an intrinsically universal tile set for any ``nice'' tile assembly system. Before proceeding, we must first define the notion of a ``nice'' tile assembly system. Let $\mathcal{T} = (T, \sigma, 2)$ be a tile assembly system, and $\vec{\alpha}$ be an assembly sequence in $\mathcal{T}$ whose result is denoted as $\alpha$. We say that $\mathcal{T}$ is \emph{locally consistent} if the following conditions hold.
\begin{enumerate}
\item \label{locally-consistent-cond-1} For all $\vec{m} \in \dom{\alpha} - \dom{\sigma}$,
$
\sum_{\vec{u} \in
\textmd{IN}^{\vec{\alpha}}(\vec{m})}{\textmd{str}_{\alpha(\vec{m})}(\vec{u})
} = 2,
$
where $\textmd{IN}^{\vec{\alpha}}\left(\vec{m}\right)$ is the set of sides on which the tile that $\vec{\alpha}$ places at location $\vec{m}$ initially binds. That is, every tile initially binds to the assembly with exactly bond strength equal to $2$ (either a single strength $2$ bond or two strength $1$ bonds).
\item For all producible assemblies $\alpha \in \prodasm{\mathcal{T}}$, $\vec{u} \in U_2$, and $\vec{m} \in \dom{\alpha}$, if $\alpha(\vec{m}+\vec{u})$ is defined, then the following condition holds:
    $$\textmd{str}_{\alpha(\vec{m})}(\vec{u}) > 0 \Rightarrow \textmd{label}_{\alpha(\vec{m})}(\vec{u}) = \textmd{label}_{\alpha(\vec{m}+\vec{u})}(-\vec{u}) \textmd{ and } \textmd{str}_{\alpha(\vec{m})}(\vec{u}) = \textmd{str}_{\alpha(\vec{m}+\vec{u})}(-\vec{u}).$$
\end{enumerate}

While condition \eqref{locally-consistent-cond-1} of the above definition is reminiscent of the first condition of local determinism \cite{SolWin07}, the second condition says that there are no (positive strength) label mismatches between abutting tiles. However, we must emphasize that a locally consistent tile assembly system need not be directed, and moreover, even a locally deterministic tile assembly system need not be locally consistent because of the lack of any kind of ``determinism restriction'' in the latter definition. Our main result is the following.

\begin{theorem}[Main theorem]
\label{maintheorem}
Let $\frakC$ be the set of all locally consistent tile assembly systems. There exists a finite tile set $U$ that is intrinsically universal for $\frakC$.
\end{theorem}

In the remainder of this section, we prove Theorem~\ref{maintheorem}, that is, we show that for every locally consistent tile assembly system $\mathcal{T} = (T,\sigma,2)$, there exists a seed assembly $\sigma_{\mathcal{T}}$, such that the tile assembly system $\mathcal{U}_{\mathcal{T}} = \left(U,\sigma_{\mathcal{T}},2\right)$ simulates $\mathcal{T}$ with a resolution loss $c \in \mathbb{N}$ that depends only on the glue complexity of~$\mathcal{T}$. Instead of giving an explicit (and tedious) definition of the tile types in $U$, we implicitly define $U$ by describing how $\mathcal{U}_{\mathcal{T}}$ simulates $\mathcal{T}$.

\subsection{High-Level Overview}

Intuitively, $\mathcal{U}$ simulates $\mathcal{T}$ by growing ``supertiles'' that correspond to tile types in $T$. In other words, every supertile is a $c \times c$ block of tiles that is mapped to a tile type $t \in T$.  To do this, each supertile that assembles in $\mathcal{U}_{\mathcal{T}}$ contains the full specification of $T$ as a lookup table (a long row of tiles that encodes all of the information in the set of tile types~$T$), analogous to the genome of an organism being fully replicated in each cell of that organism, no matter how specialized the function of the cell.  This lookup table is carefully propagated through each supertile in $\mathcal{U}_{\mathcal{T}}$ via a series of ``rotation'' and ``copy'' operations -- both of which are well-known self-assembly primitives.

In the table, we represent each (glue,direction) pair as a binary string, and represent the tile set as a table mapping 1-2 input glue(s) to 0-3 output glue(s). Since each tile type of $\mathcal{T}$ may not have well-defined input sides, when two supertiles representing tiles of $\mathcal{T}$ must potentially cooperate to place a new supertile within a block adjacent to both of them, it is imperative that each grows into the block in such a way as to remain unobtrusive to the other supertile. This is done with a ``probe'' that grows toward the center of the block, as shown in Figure \ref{fig:supertile-NS-detailed}. At the moment the probes meet in the middle, they ``find out'' in what direction the other input supertile lies, and at that point decide in which direction to grow the rest of the forming supertile. so as to avoid the tiles that were already placed as part of the probes. We do not know how to deal with three probes at once, which is the reason both parts of the definition of locally consistent, which imply that only two input probes will ever be present at one time. The next step is to bring the values of two input glues together before doing a lookup on the table, because they are both needed to simulate cooperation. The table must be read and copied at the same time, otherwise the planarity of the tiles would hide the table as it is read and it could not be propagated to the output supertiles. Many choices made in the construction, such as the relative positioning of glues/table, or the counter-clockwise order of assembly, are choices that simply were convenient and seemed to work, but are not necessarily required.

\subsection{Construction of the Lookup Table}

In order to simulate the behavior of $\mathcal{T}$ with $\mathcal{U}_{\mathcal{T}}$, we must first encode the definition of $T$ using tiles from $U$. We will do this by constructing a ``glue lookup table,'' denoted as $\mathbf{T}_\calT$, and is essentially the self-assembly version of a kind of hash table. Informally, $\mathbf{T}_\calT$ is a (very) long string (of tiles from $U$) consisting of two copies of the definition of the tile set $T$ separated by a small group of spacer symbols.  The left copy of the lookup table is the \emph{reverse} of the right copy.  The lookup table maps all possible sets of input sides for each tile type $t \in T$ to the corresponding sets of output sides.

\subsubsection{Addresses}
The lookup table $\mathbf{T}_\calT$ consists of a contiguous sequence of ``addresses,'' which are formed from the definition of $T$. Namely, for each tile type $t \in T$, we create a unique binary key for each combination of sides of $t$ whose glue strengths sum to exactly $2$.  Each of these combinations represents a set of sides which \emph{could} potentially serve as the input sides for a tile of type $t$ in a producible assembly in $\mathcal{T}$.

We say that a {\it pad} is an ordered triple $(g, d, s)$ where $g$ is a glue label in $T$, $d \in \{N, S, E, W\}$ is an edge direction, and $s \in \{0, 1, 2\}$ is an allowable glue strength.  Note that a set of four pads -- one for each direction $d$ -- fully specifies a tile type.  We use $\textmd{Pad}(t,d)$ to denote the pad on side $d$ of the tile type $t \in T$

Let $\textmd{Bin}(p)$ be the binary encoding of a pad $p = (g, d, s)$, consisting of the concatenation of the following component binary strings:
\begin{enumerate}
\item{$g$ (\emph{glue specification}):  Let $G$ be the set of glue types from all edges with positive glue strengths in $T \cup \left\{g_{\textmd{null}}\right\}$ (a.k.a., the null glue). Fix some ordering $g_{\textmd{null}} \leq g_0 \leq g_1 \leq \cdots$ of the set $G$.  The binary representation of $g_i$ is the binary value of $i$ padded with $0$'s to the left (as necessary) to ensure that the string is exactly $\lceil \log(|G|+1) \rceil$ bits.}
\item{$d$ (\emph{direction}):  If $d = N$ ($E$, $S$, or $W$), append $00$ ($01$, $10$, or $11$, respectively).}
\item{$s$ (\emph{strength}):  If $s = 1$ ($2$) append $0$ ($1$).}
\end{enumerate}

Note that $\lceil \log(|G|+1) \rceil + 2 + 1$ is the length of the binary string encoding an arbitrary pad $p$, and is a constant that depends only on $T$.

An {\it address} is a binary string that represents a set of pads which, themselves, can potentially serve as the input sides of some tile type $t \in T$. It can be composed of one of the two following binary strings:
\begin{enumerate}
\item{A prefix of zeros, $0^{\left\lceil \log(|G|+1) \right\rceil + 3}$, followed by $\textmd{Bin}(p)$ for $p = (g, d, 2)$, or}
\item{the concatenation of $\textmd{Bin}(p_1)$ and $\textmd{Bin}(p_2)$ for $p_1 = (g_1, d_1, 1)$ and $p_2 = (g_2, d_2, 1)$.  The ordering of $\textmd{Bin}(p_1)$ and $\textmd{Bin}(p_2)$ in an address must be consistent with the following orderings: $EN, SE, WS, NW, NS, EW$.}
\end{enumerate}

Note that it is possible for more than one tile type $t \in T$ to share a set of input pads and therefore an address.

\subsubsection{Encoding of $T$}

We will now construct the string $w_\calT$, which will represent the definition of $T$. Intuitively, $w_\calT$ will be composed of a series of ``entries.'' Each entry is associated to exactly one address of a tile type $t \in T$ and specifies the pads for the output sides of $t$.  In this way, once the input sides for a supertile have formed, the corresponding pads can be used to form an address specifying (a set of) appropriate output pads. Note that since more than one tile type may share an address in a nondeterministic tile assembly system, more than one tile type may share a single entry.

We define an {\it entry} to be a string beginning with `\#' followed by zero or more ``sub-entries'', each corresponding to a different tile type, separated by semicolons.  Let $A$ be the set of all binary strings representing every address created for each $t \in T$. The string $w_\calT$ will consist of $1+\max A$ entries for addresses $0$ to $\max A$.  The $i^{\textmd{th}}$ entry, denoted as $e_i$, corresponds to the $i^{\textmd{th}}$ address, which may or may not be in $A$ (if it is not, then $e_i$ is empty).

We say that a {\it sub-entry} consists of a string specifying the pads for the output sides of a tile type $t \in T$. Let $e_i$ be the entry containing a given sub-entry (note that $i$ is the address of $e_i$), and $T_i \subseteq T$ be the set of tile types addressable by $i$ (i.e., the set of tile types for which $i$ is a valid address).  The entry $e_i$ will be comprised of exactly $\left|T_i\right|$ sub-entries.  For $0 \leq k < j$, the $k^{\textmd{th}}$ sub-entry in $e_i$, where $t_k \in T_i$ is the $k^{\textmd{th}}$ element of $T_i$ (relative to some fixed ordering), is the string $\textmd{OUT}(N),\textmd{OUT}(E),\textmd{OUT}(S),\textmd{OUT}(W)$ (the commas in the previous string are literal) with $\textmd{OUT}(d) = \textmd{Bin}(\textmd{Pad}(t_k,d))^R$ if the glue for $\textmd{Pad}(t_k,d)$ is not $g_{\textmd{null}}$ and $d$ is not a component of the address $i$, otherwise $\textmd{OUT}(d) = \lambda$.  Intuitively, a sub-entry is a comma-separated list of the (reversed) binary representations of the pads for an addressed tile type, but including only pads whose glues are not $g_{\textmd{null}}$ and whose directions are not a part of the address (and therefore input sides). We will now use the string $w_\calT$ to construct the lookup table $\mathbf{T}_\calT$.

\subsubsection{Full specification of $\mathbf{T}_{T}$}
We now give the full specification for the lookup table $\mathbf{T}_\calT$. First, define the following strings: $w_0 =$ `$>$', $w_1 =$  `$<\%\%>$', $w_2=$ `$<$'. Now let $\mathbf{T}_\calT$ be as follows:
$
    \mathbf{T}_\calT = \mathrm{sb}(w_0 \circ w_\calT \circ w_1 \circ (w_\calT)^R \circ w_2),
$
where, for strings $x$ and $y$, $x \circ y$ is the concatenation of $x$ and $y$, and $\mathrm{sb}:\Sigma^* \to \Sigma^*$ is defined to ``splice blanks'' into its input: between every pair of adjacent symbols in the string $x$, a single `$\tmblank$' (blank) symbol is inserted to create $\mathrm{sb}(x)$. This splicing of blanks is required to be able to read from the table without ``locking it from view'', when reading the table for operations that require growing a column of tiles in towards the table (as opposed to away from it), a blank column is used, and for growing a column away from the table, a symbol column is used so that the symbol can be propagated to the top of the column for later copying.

\subsubsection{The Lookup Procedure}

In our construction, when a supertile $t^*$ that is simulating a tile type $t \in T$ forms, we must overcome the following problem: once we combine the input pads (given as the output pads of the supertiles to which $t^*$ attaches), how do we use $\mathbf{T}_\calT$ to lookup the output pads for $t^*$? In what follows, we briefly describe how we achieve this. In other words, we show how an address, a string of random bits, and a copy of $\mathbf{T}_\calT$ are used to compute the pad values for the non-input sides of a supertile.  A detailed figure and example of this procedure can be found in the technical appendix.

For ease of discussion and without loss of generality, we assume that the row of tiles encoding \textbf{T}$_{\mathcal{T}}$ (assembled West to East) and the column of tiles encoding an address and a random string of bits (assembled North to South at the West end of \textbf{T}$_{\mathcal{T}}$) are fully assembled, forming an `L' shape with no tiles in the area between them.  For other orientations of the table and address the logical behavior is identical, simply rotated.

Intuitively, the assembly of the lookup procedure assembles column wise in a zig-zag fashion from left to the right.  In the ``first phase,'' a counter initialized to $0$ is incremented in each column where the value of the tile in the representation of $\mathbf{T}_\calT$ is a `;', thus counting up at each entry contained in $\mathbf{T}_\calT$.  Once that number matches the value of the given address (which, along with the random bits is copied through this procedure), the entry $e$ corresponding to that address has been reached and a new counter begins which counts the number of sub-entries $n$ in that entry. Note that for directed tile systems, $n \leq 1$. Once the end of that entry is encountered, yet another counter, initialized to $0$, begins and increments on each remaining entry until the end of the first copy of $w_\calT$ is reached (the number $n$ is propagated to the right).  This counts the number of entries, denoted as $m$, between $e$ and the end of the lookup table.  The ``second phase'' is used to perform, in some sense, an operation equivalent to calculating $p = b \mod n$, where $b$ is the binary value of the string of random bits required for the lookup procedure (this is how we simulate nondeterministic assemblies).  This selects the index of the sub-entry in $e$ which will be used, completing the random selection of one of the possibly many tile types contained in entry $e$.

In the current version of our construction, we merely use a random number selection procedure reminiscent of the more involved (but more uniform) random selection procedures discussed in \cite{RNSSA}. Although it is possible to incorporate these more advanced techniques into our construction (and thus achieve a higher degree of uniformity in the simulation of randomized tile systems), we choose not to do so for the sake of simplicity.

Next, a reverse counter, a.k.a., a subtractor, counts down at each entry from $m$ to $0$, and by the way we constructed $\mathbf{T}_\calT$, this final counter obtains the value $0$ at the entry $e$ (in the reverse of $w_\calT$).  Now, another subtractor counts from $p$ to $0$ to locate the correct sub-entry that was selected randomly.  Finally, each pad in the sub-entry is rotated ``up and to the right,'' and the group of pads is propagated through the remainder of the lookup table, thus ending with the values of the non-input pads represented in the rightmost column.

\subsection{Supertile design}

A supertile $s$ is a subassembly in $\mathcal{U}_{\mathcal{T}}$ consisting of a $c \times c$ block of tiles from $T$, where~$c$ depends on the glue complexity of $T$.  Each $s$ can be mapped to a unique tile type $t \in T$. In our construction there are two logical supertile designs.  The first, denoted \emph{type-0}, simulates tile additions in $\mathcal{T}$ in which there are $2$ input sides, each with glue strength $= 1$.  The second, denoted \emph{type-1}, simulates  the addition of tiles via a single strength 2 bond.

While there are several differences in the designs of \emph{type-0} and \emph{type-1} supertiles, one commonality is how their edges are defined.  Namely each input or output edge of any supertile is defined by the same sequence of variable values. Since the edges for each direction are rotations of each other, we will discuss only the layout of the south side of a supertile.  From left to right, the tiles along the south edge of a supertile will represent a string formed by the concatenation (in order) of the strings:  $\mathbf{T}_\calT$, $\textmd{Bin}(\textmd{Pad}(t,S))$, $0^{c'}$, $\textmd{Bin}(\textmd{Pad}(t,S))$, and $\mathbf{T}_\calT$. Note that $c'$ is a constant that depends on the glue complexity~of~$T$.

\subsubsection{Type-0 Supertiles (i.e., simulating tiles that attach via two single-strength bonds)}
\begin{figure}[htp]
\begin{center}
\includegraphics[width=4.5in]{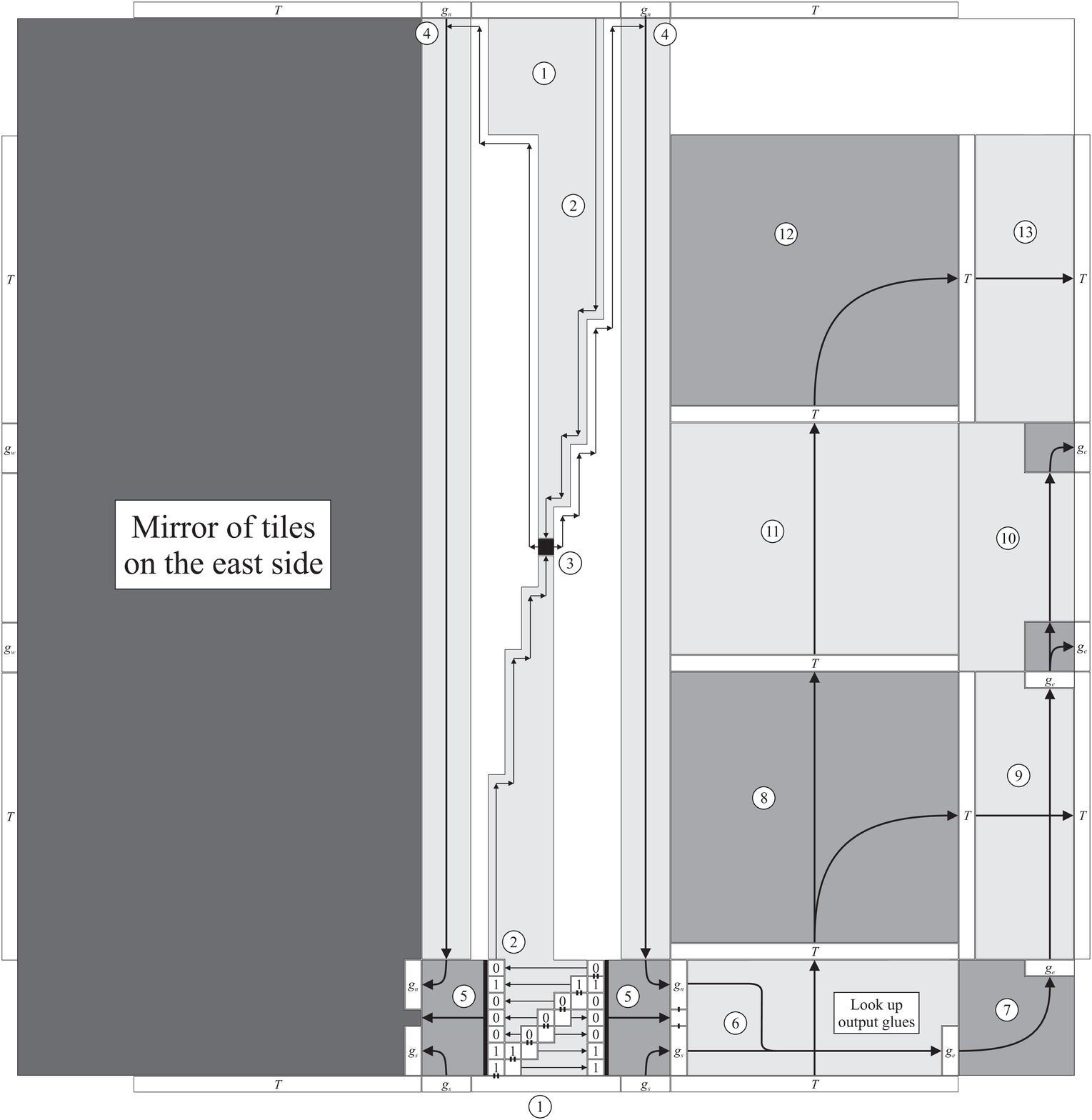} \caption{\label{fig:supertile-NS-detailed} \small NS supertile}
\end{center}
\end{figure}
When a tile binds to an assembly in $\mathcal{T}$ with two input sides whose glues are each single strength, there are ${4 \choose 2} = 6$ possible combinations of directions for those input sides: north and east (NE), north and south (NS), north and west (NW), east and south (ES), east and west (EW), and south and west (SW).  These combinations can be divided into two categories, those in which the sides are opposite each other (NS and EW), and those in which the sides are adjacent to each other (NE, NW, ES, and SW).

\textbf{Opposite Input Sides:} Supertiles which represent tile additions with two opposite input sides, NS and EW, are logically identical to rotations of each other, so here we will only describe the details of a supertile with NS input sides.
\begin{figure}[htp]
\begin{center}
    \subfloat[][]{%
        \label{fig:st0}%
        \includegraphics[width=1.0in]{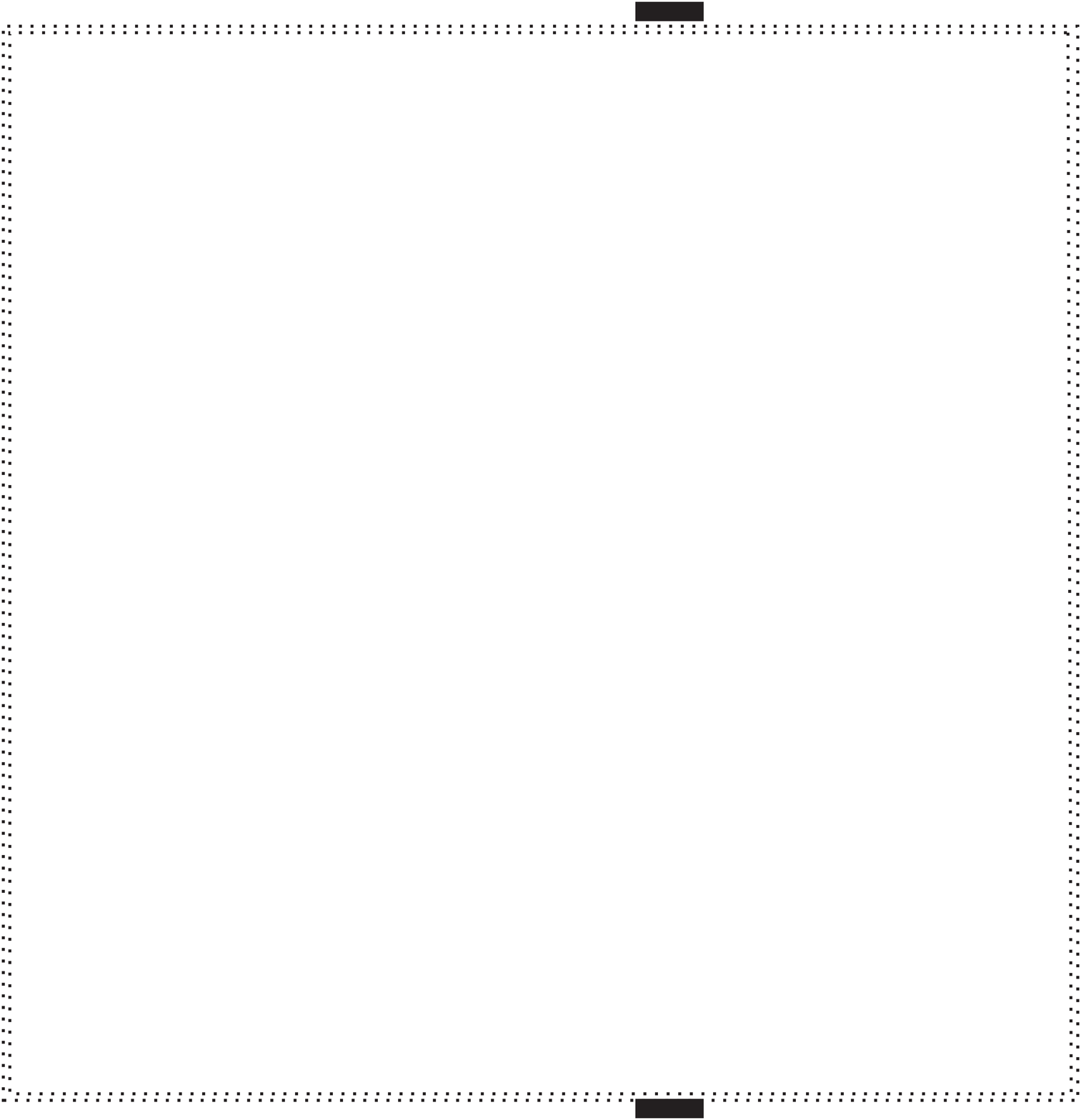}} \hspace{10pt}
        \subfloat[][]{%
        \label{fig:st1}%
        \includegraphics[width=1.0in]{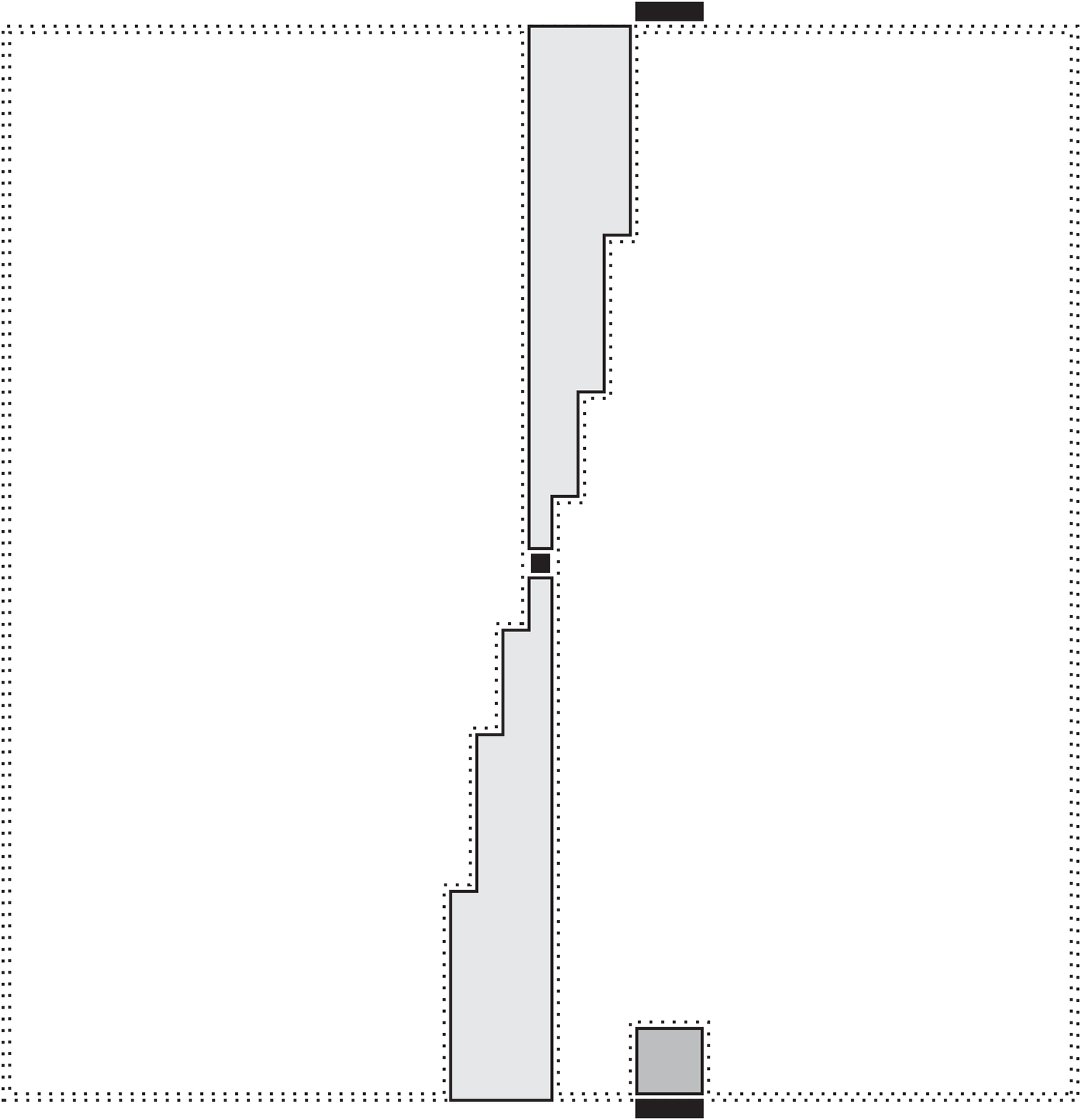}} \hspace{10pt}
        \subfloat[][]{%
        \label{fig:st2}%
        \includegraphics[width=1.0in]{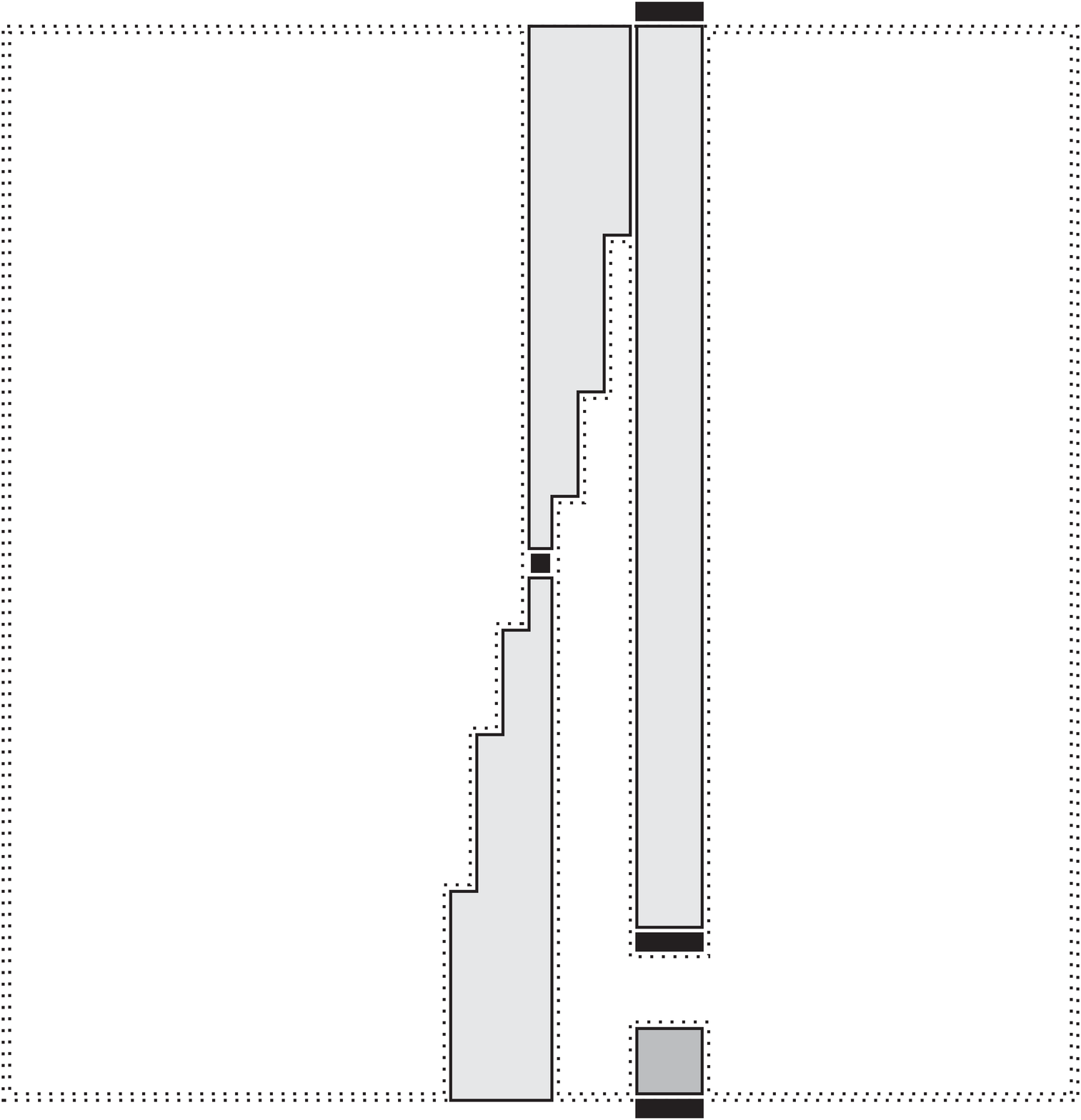}} \hspace{10pt}
        \subfloat[][]{%
        \label{fig:st3}%
        \includegraphics[width=1.0in]{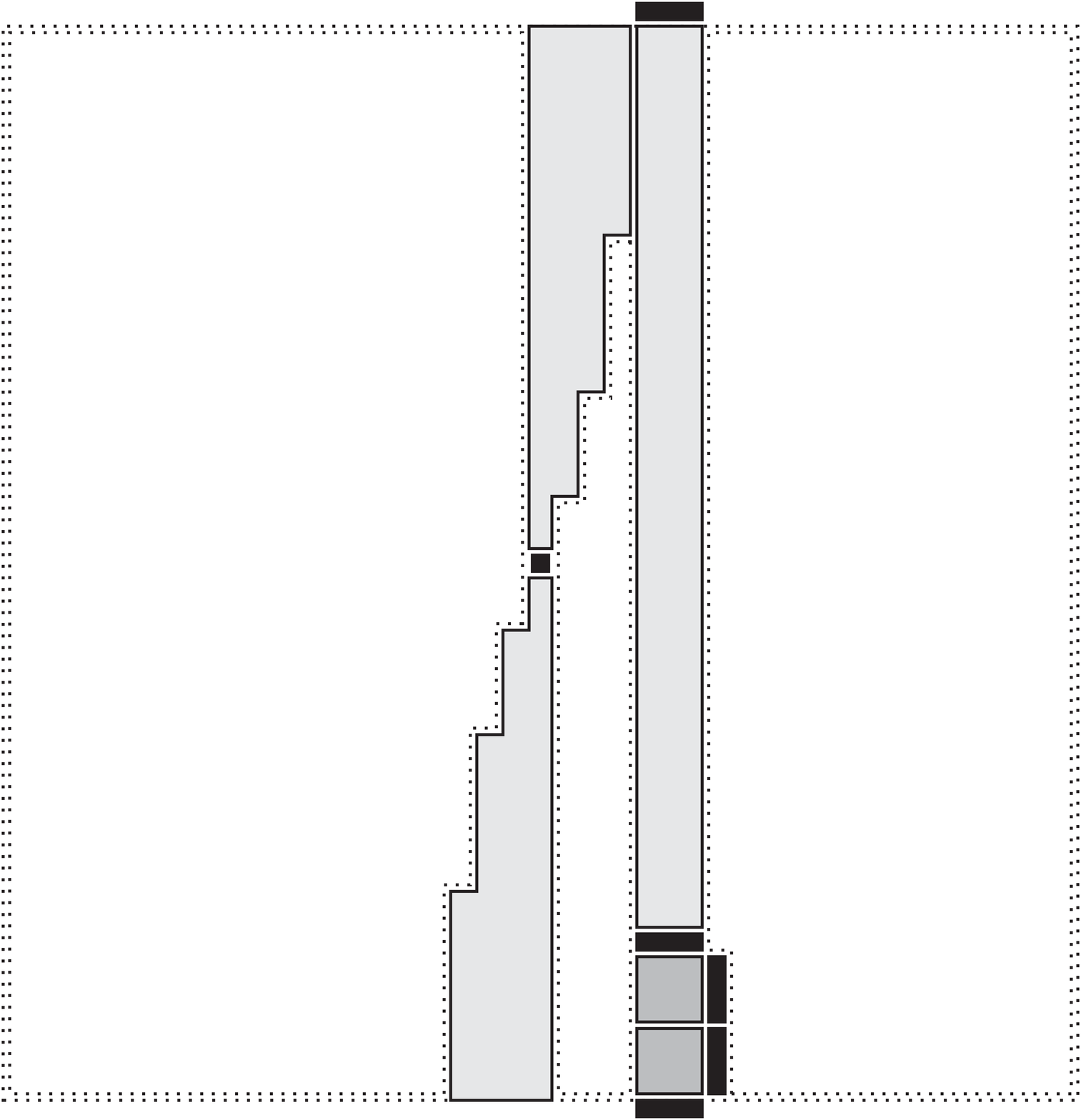}}
        \\
        \subfloat[][]{%
        \label{fig:st4}%
        \includegraphics[width=1.0in]{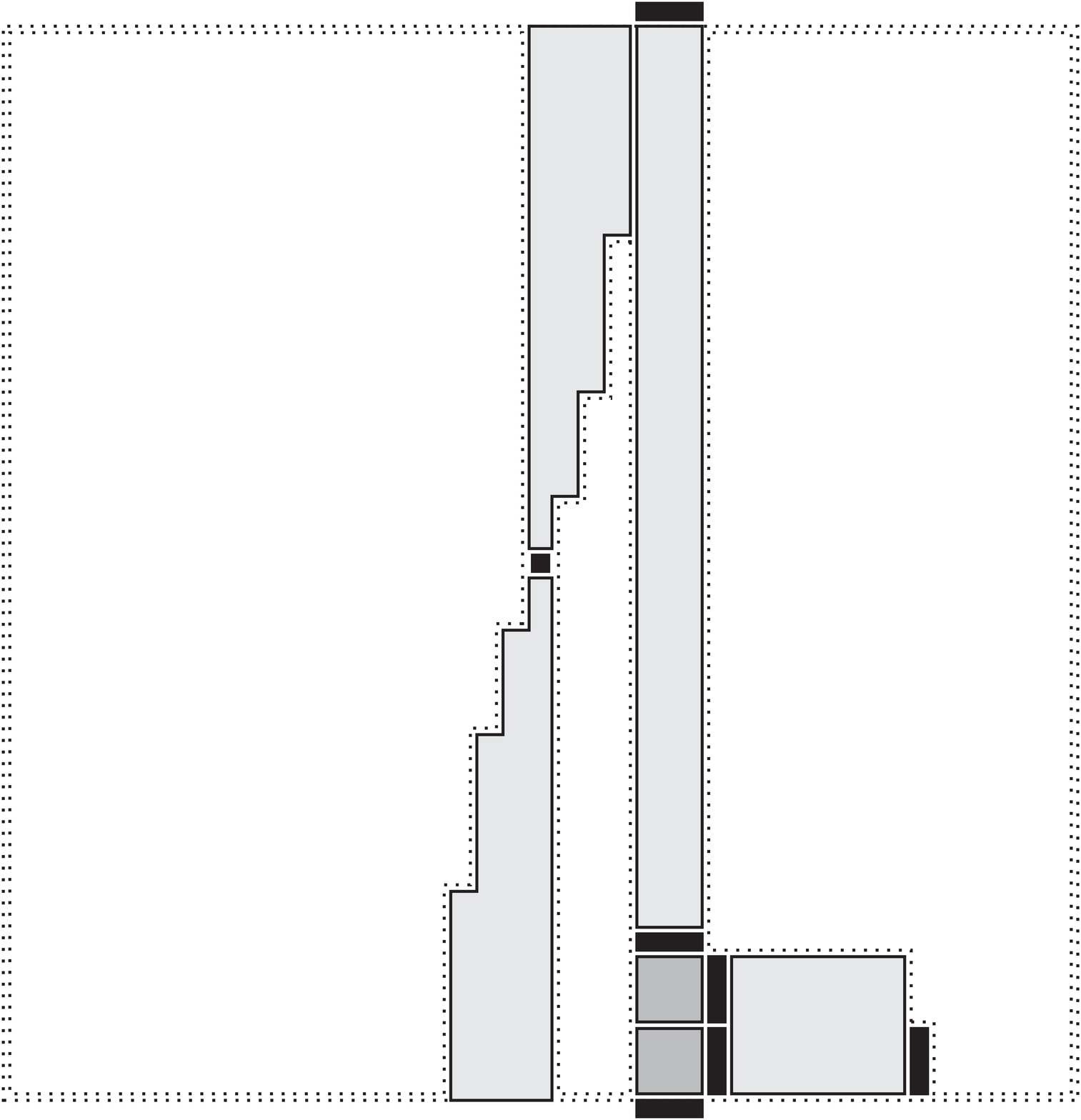}} \hspace{10pt}
        \subfloat[][]{%
        \label{fig:st5}%
        \includegraphics[width=1.0in]{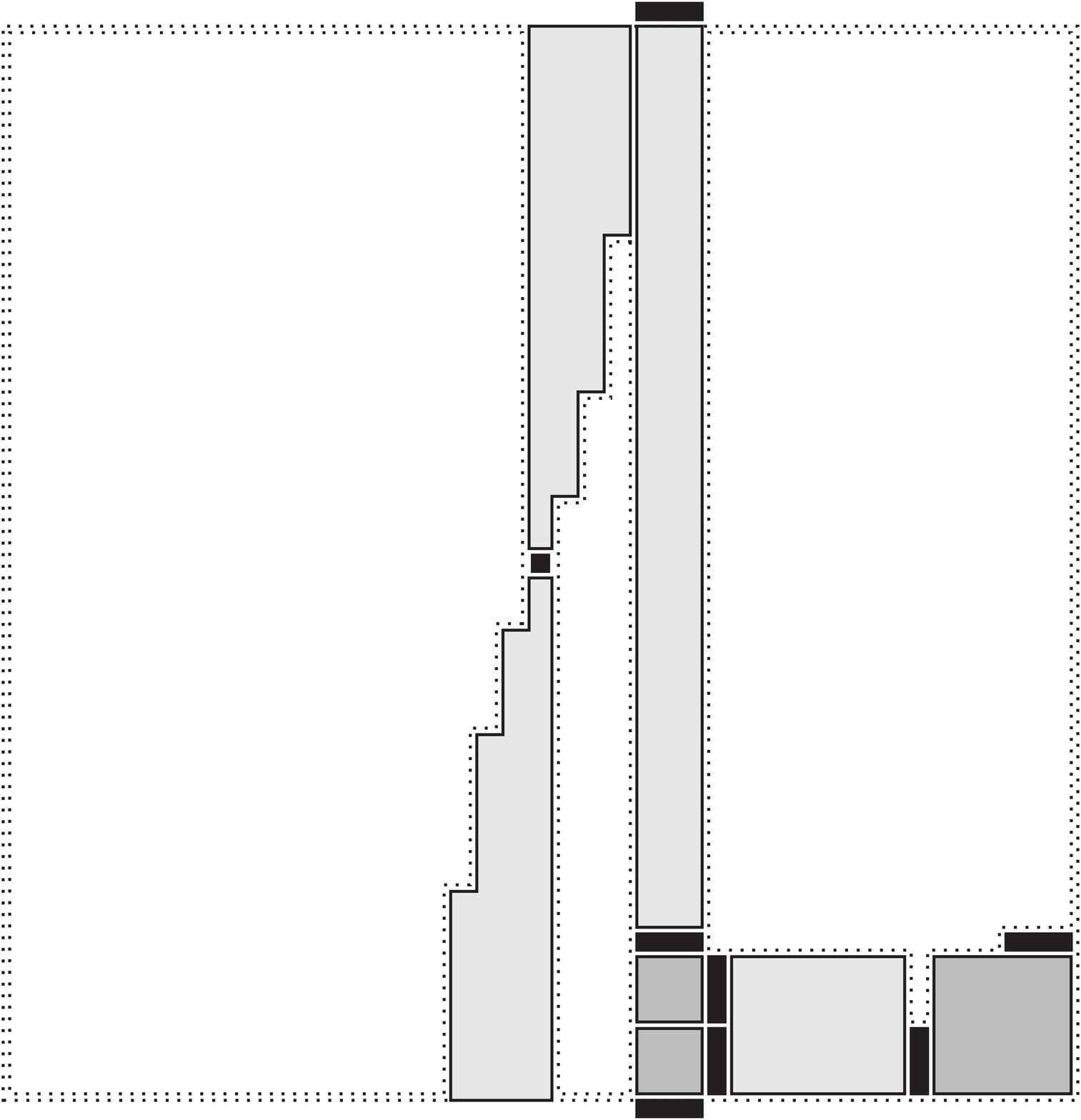}} \hspace{10pt}
        \subfloat[][]{%
        \label{fig:st6}%
        \includegraphics[width=1.0in]{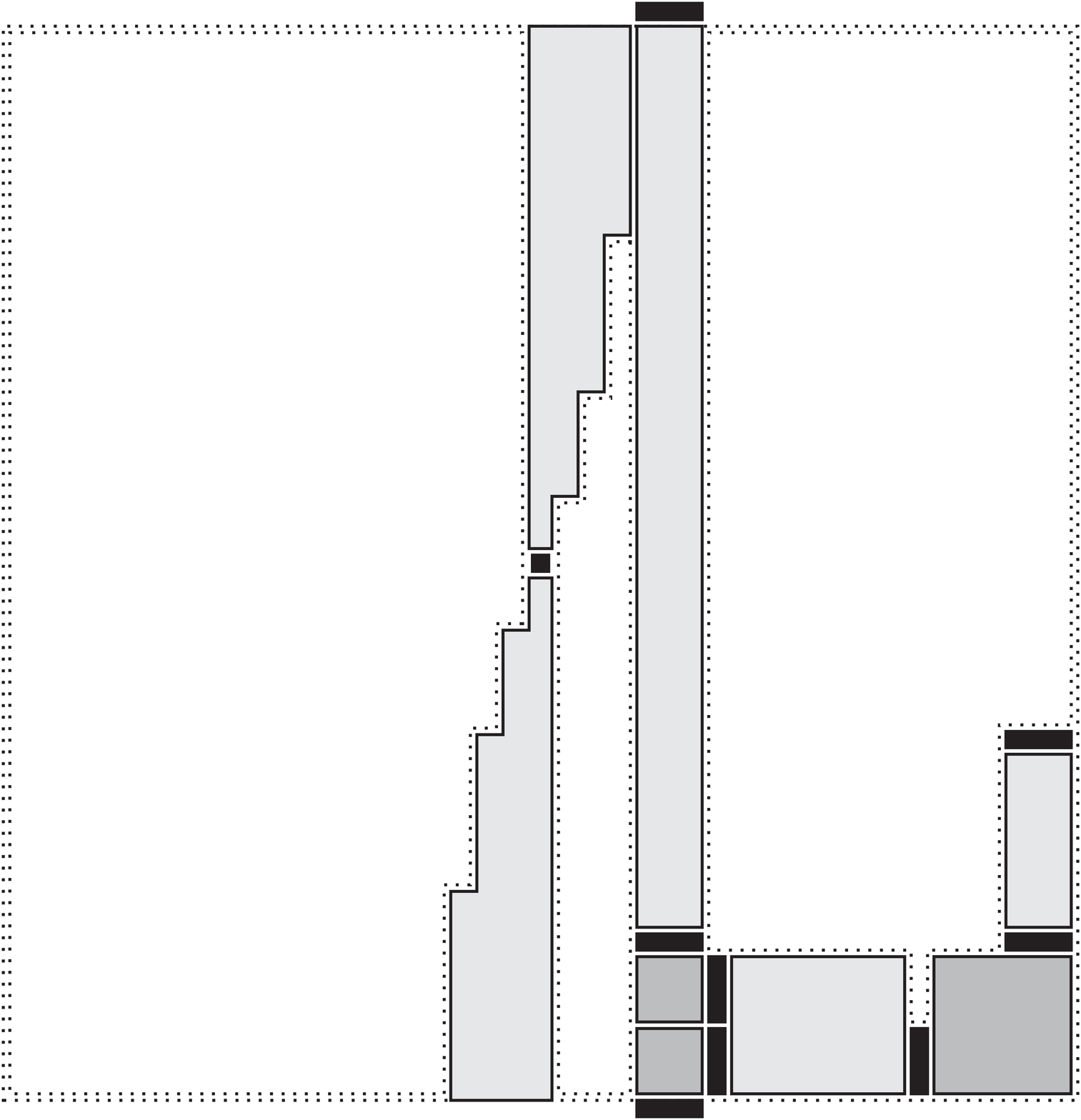}} \hspace{10pt}
        \subfloat[][]{%
        \label{fig:st7}%
        \includegraphics[width=1.0in]{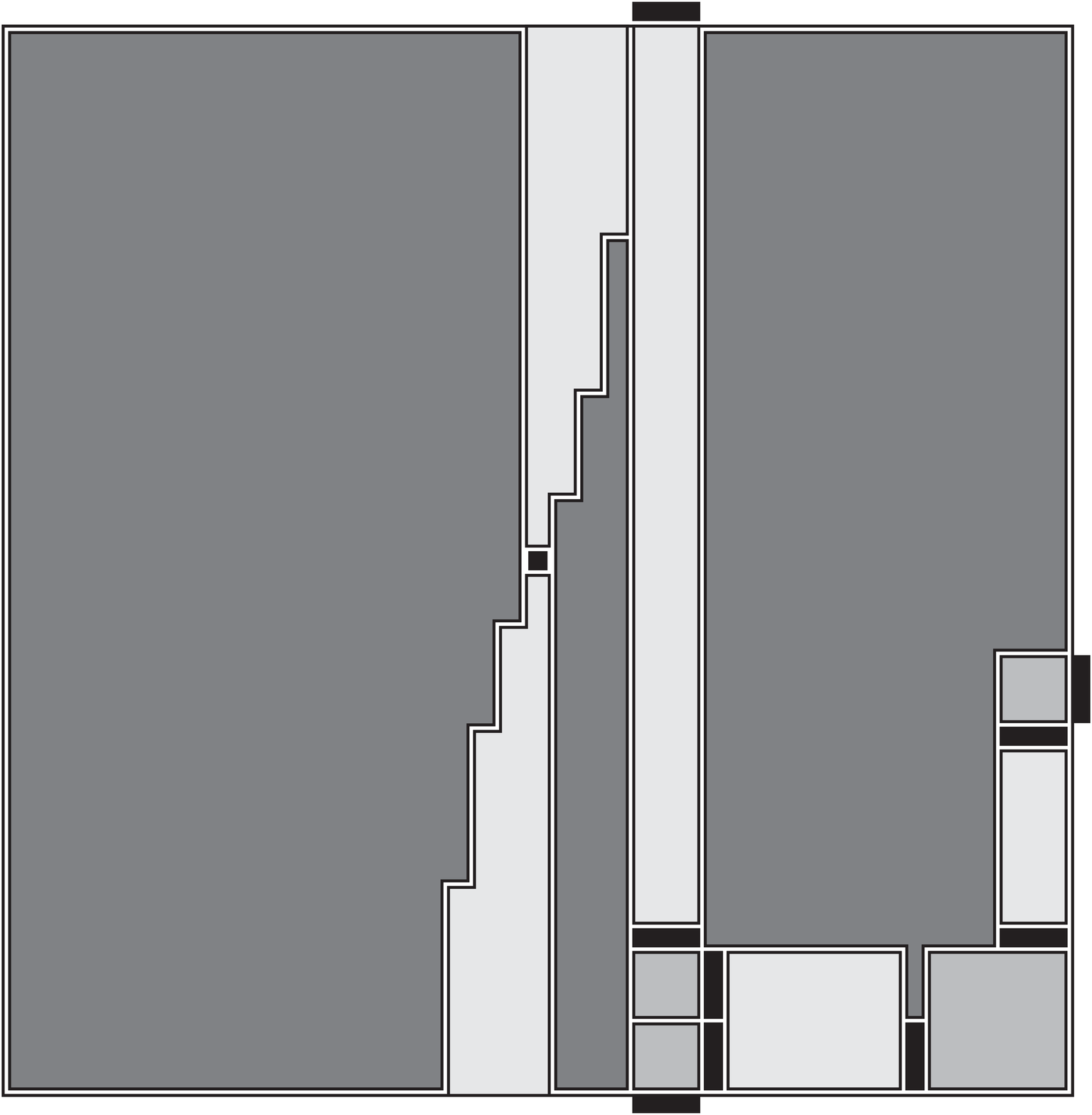}}
    \caption{\label{fig:st} \small Intuitive depiction of (a portion of) the self-assembly of a type-0 supertile. Note that the lookup procedure is performed in \subref{fig:st3} and \subref{fig:st4}.}
\end{center}
\end{figure}
Figure~\ref{fig:supertile-NS-detailed} shows a detailed image depicting the formation of an NS supertile, with arrows giving the direction of growth for each portion and numbers specifying the order of growth.  For ease of discussion and without loss of generality, we assume that the rows of tiles which form the input sides of a supertile have fully formed before any other part of the supertile assembles.  The first portions to assemble are the center blocks to the interior of each input side, labeled 1.  This subassembly forms a square in which a series of nondeterministic selections of tile types is used to generate a random sequence of bits.  These bits are propagated to the left and right sides of the block, to ensure that each side uses the same random bits for the randomized selection after the sides have been ``sealed off'' from each other by ``probes'' described next.  Once that block has completed, a $\log$-width binary subtractor, which is half the width of the block, assembles.  The subtractors from the north and south count down from a specified value (that depends on $\mathcal{T}$ and is encoded into the seed supertile) to $0$, and shrink in width until they terminate at positions adjacent to the center square of the block.  These subtractors are ``probes'' that grow to the center where the direction of the input sides (the \emph{type}) is detected.  It is at this point that the central (black in the figure) tile can attach.  It is this tile which determines the \emph{type} of the supertile (NS in this case) because it is unique to the combination of directions from which the inputs came.  At this point, symmetry is broken and two paths of tiles assemble from the center back towards the north side.  They in turn initiate the growth of subassemblies which propagate the value of the north input pad down towards the South of the supertile.  Once that growth nears the southern side, the two input pads are rotated and brought together, with this combination of input pads forming an \emph{address} in the lookup table.  In the manner described previously, this address along with the random bits generated within block 1 (which are also passed through block 5) is used to form the subassembly of block 6 whose southern row contains a representation of $\mathbf{T}_\calT$ and results in the correct output pads being represented in the final column of that block.  Note that Figure~\ref{fig:supertile-NS-detailed} only shows the details of the east side of the block since the West side is an identical but rotated version.  Finally, subassemblies 7 through 13 form which rotate and pass the necessary information to the locations where it must be correctly deposited to form the output sides of the supertile.  Every side of a supertile that is not an input side receives an output pad, even if it is for the null glue (in which case it does not initiate the growth of the input side of a possible adjacent supertile).